\documentstyle[aps,preprint,prb]{revtex}

\renewcommand{\theequation}{\arabic{section}.\arabic{equation}}

\def\dd{{\it\Delta}}

\def\eps{{\it\epsilon}}
\def\om{{\it\omega}}
\def\gam{{\it\gamma}}
\def\Gam{{\it\Gamma}}
\def\ba{\begin{eqnarray}}
\def\ea{\end{eqnarray}}
\def\nn{\nonumber}

\def\e{\mbox{e}}
\def\d{\mbox{d}}

\def\<{\stackrel{<}{\sim}}
\def\>{\stackrel{>}{\sim}}

\begin{document}

\title{Relaxation Dynamics of Multi-Level Tunneling
 Systems\footnote{Appears in the Journal of Chemical Physics.}}

\author{Peter Neu$^{1)}$  and Andreas Heuer$^{2)}$}

\maketitle
\thispagestyle{empty}
\begin{center}
 {\it $^{1)}$Department of Chemistry
 \\ \qquad Massachusetts Institute of Technology, Cambridge, MA.
02139}\\[1cm]
{\it $^{2)}$Max-Planck Institut f\"ur Polymerforschung, \\
Ackermannweg 10, D-55128 Mainz, Germany}
\end{center}
%\date{\today}

% \vspace{5cm}

\begin{abstract}
A quantum mechanical treatment of an asymmetric double-well potential (DWP)
interacting with a heat bath is presented for circumstances where
the  contribution of higher vibrational levels to the relaxation  dynamics
cannot be excluded from consideration. The deep quantum limit
characterized by a discrete energy spectrum near the barrier top is considered.
The investigation is motivated by simulations  on a computer glass
which show that the considered parameter regime is ``typical'' for DWPs
being responsible for the relaxation peak of sound absorption in glasses.
Relaxation dynamics  resembling  the spatial-
and energy-diffusion-controlled limit
of the classical Kramers' problem,  and
Arrhenius-like  behavior is found under specific conditions.\\

%\noindent PACS-number(s): 61.40, 63.50, 77.40
 \end{abstract}

%\pacs{PACS numbers: 61.40, 63.50, 77.40}

\newpage

\setcounter{page}{1}

%\narrowtext

\section{Introduction}
Relaxational dynamics of materials which evolve under slow
structural change at low temperature can frequently be studied
in a double-well potential (DWP) picture that allows tunneling between
the two wells. A prototype of such a system, important in many
chemical and biological systems, is  the transfer
of hydrogen atoms along preexisting hydogen bonds for instance  in
crystalline benzoic acid   and
carboxyl dimers. \cite{Na,Sk,ME,SS,HH,Mak} Another example is the tunneling
of SiO$_4$- or GeO$_4$-tetraeder in amorphous SiO$_2$ and GeO$_2$.\cite{Hunk,Rau}

One of the key issues in understanding the relaxation process
is the coupling of the motion of the light tunneling particle
to the havier surrounding atoms constituting the heat bath.
In a condensed phase the coupling leads to structural
rearrangements of the environment which provides a mechanism
of relaxation for the system.

The theoretical description of dissipative DWP-dynamics
 has been widely elaborated  in the low temperature limit where
 a two-level description (spin-boson model) 
is sufficient. \cite{Legg,Weisb,Var,Mode} At higher  temperatures  where
the excitation of vibrational levels can no longer be neglected,
 the effect of intra-well relaxation has been given explicitly
by  Silbey and coworkers \cite{PS}
for a symmetric DWP. Lateron Meyer and Ernst \cite{ME}
considered  a biased DWP  with  crystalline benzoic acid dimer as an example.
In both papers the investigations have been restricted to  a regime
where the tunneling splitting $\dd_n$ of all relevant  doublets is still
very much less than the interdoublet spacing, i.e., $\dd_n\ll\om_0$
where  $\om_0$ is the frequency of small oscillations around the two minima
of the DWP.
They found in the limit of fast intra-well vibrational relaxation --
where the population of the doublets is in thermal equilibrium --
 that  the escape rate is the
thermally averaged tunneling rate into the lower well,
\begin{equation}\label{2}
k_t = {\textstyle {1\over 2}} \langle \Gam\rangle_\beta\ .
\end{equation}
In many practical applications, however,
  whenever the two-state approximation seemed no longer sufficient,
  a simple classical relaxation over the barrier $V$  with an Arrhenius rate
 \ba\label{1}
k_{\rm cl} \ = \ {\textstyle \frac{\om_0}{2\pi}} \  \e^{-V/k_{\rm B} T}
\ea
has been used successfully.\cite{Sk,Hunk,Rau}

Hence, concerning the understanding of relaxation dynamics
in many chemical and physical systems at
higher temperatures, two
questions are of major
importance: (i) How dense are the energy levels of the DWP as compared
to the potential height, i.e., is the classical picture underlying
Eq. (\ref{1}) justified, or does the relaxation dynamics depend on the
quantum mechanical eigenvalues of the DWP? (ii) Is the DWP coupled to
bath modes with frequencies of the order of its potential height, and
does the relaxation dynamics depend on the parameters of the heat bath?

In this work these questions are analyzed for the
specific example of  DWPs in a structural glass. There
DWPs correspond to the motion of a local group of atoms or molecules
between two local energy minima. Low barrier DWPs
are known to be the source for the tunneling properties observed in
glasses like SiO$_2$ below 10~K.\cite{Phi87}
In contrast, DWPs with higher barriers
(in SiO$_2$: $\langle V_{\rm peak} \rangle/k_{\rm B} \approx 500$ K)
are believed to be responsible for the
relaxation peak of sound absorption, observed around 50 K in SiO$_2$.\cite{Hunk,Rau}
Our goal is to answer the above questions for these DWPs in
order to obtain a closer understanding of the relaxation dynamics
at the absorption peak.

To this aim we first perform in Sec. 2  computer simulations on a model glass
to determine the eigenvalue spectrum and the interaction with phonons
of those  DWPs which are responsible for the relaxation dynamics
in glasses at higher temperatures. We find that DWPs relevant at the sound
absorption peak
only {\it contain a few energy levels below the their barrier height} $V$
and are {\it coupled to bath modes with frequencies as large as} $V$.
This serves us as a motivation to analyse the relaxation dynamics in a {\it
single}
DWP as depicted in Fig. 1 which is  characterized by the parameter regime
 \begin{equation}\label{TR}
 k_{\rm B} T \< \hbar\om_0 \< V\ ,
 \end{equation}
for instance,  $T\approx 30 - 50$~K, $V/k_{\rm B}\approx 300$~K and
$V/\hbar\omega_0\approx 2$.
In Sec. 3 a simple model is presented which allows an
analytical solution of the relaxation dynamics by applying standard
approximation, as shown in Sec. 4.
With this model we can discuss  in Sec. 5 the apparent paradox
that a classical Arrhenius rate does well
in many practical applications although the relaxation dynamics
is far from being classical. In Sec. 6 we close with a short summary.

\setcounter{equation}{0}
\section{Simulations}
In previous work Heuer and Silbey have developed an algorithm
which  systematically searches for DWPs in glasses simulated on
the computer.\cite{Heu1,Heu2} This algorithm has been readily
applied to the specific case
of NiP which can be described as a binary Lennard-Jones-type glass.
For general reasons one expects that $V_{\rm peak}$ is approximately
proportional to the glass transition temperature.\cite{Heu2} Hence
from $T_g$(NiP)/$T_g$(SiO$_2) \approx 2/3$  one may roughly
estimate $V_{\rm peak}$ (NiP)$/k_{\rm B}$ $\approx$ 300~K. We proceed in two
steps.
First we present a rough estimation of the number of energy levels
in DWPs with $V \approx V_{\rm peak}$. Second we explicitly
determine for the NiP computer glass
the interaction of the DWP with the heat bath.

For the estimation of the number of energy levels below the
potential height it is sufficient to consider a symmetric DWP.
The potential energy along the reaction coordinate $x$, which
is defined as the collective coordinate connecting the two
local energy minima,
can be expressed
as a quartic polynomial
\begin{equation} \label{pot}
V(x) = \frac{16 V}{d^4}(x - d/2)^2 (x + d/2)^2
\end{equation}
 where $V$ is the potential height and $d$ the distance
of both minima.
We chose the DWP symmetric around $x = 0$.
The length scale $d$ is defined such that the collective
dynamics along this DWP can be viewed as the dynamics of a single
particle with mass $m$ where $m$ is the average mass of the
atoms (for NiP $m = 56 m_p$ with $m_p$ the proton mass).

It is straightforward to calculate the vibrational frequency
$\omega_0$ in both wells as
\begin{equation}
\label{eqdwp1}
\omega_0^2 = \frac{32 V}{m d^2}.
\end{equation}
In Ref.  \cite{Heu2} it has been shown that for NiP the value of the
prefactor $16V/d^4$ of the quartic term
for symmetric DWPs
on average is given by $B A_4 / a^4$ with $B/k_{\rm B} = 4.3\times 10^4$~K,
$A_4 = 0.39$ and $a = 2.2 \times  10^{-10}$m.
Elimination of $d$ in Eq. (\ref{eqdwp1}) yields
$\omega_0^2 = 8 \sqrt{V B A_4}/(m a^2)$.
Inserting the numbers for NiP given above yields
 for $V/k_{\rm B} = 300$~K that
\begin{equation}
\label{eq55}
V / \hbar \omega_0 \approx 5.5\ .
\end{equation}
 This value should be viewed
as an upper limit since for DWPs with large barrier heights
the quartic contribution is expected to be even larger
than the average contribution. However, this would only
reduce the number of energy levels below the barrier height.

For evaluating the coupling to the
bath modes we analyzed a NiP computer glass with $N = 500$ atoms
using periodic boundary conditions. The simulations have been performed 
at zero temperature; 
details can be found in Refs. \CITE{Heu1,Heu2}.
 To first approximation the energy surface of the computer
glass adjacent
to the DWP can be written as
\begin{equation}
H = H_0(x) + x \sum_j \lambda_j y_j + {\textstyle {1\over 2}} m \sum_j
\omega_j^2 y_j^2
\end{equation}
where $H_0(x)$ describes the relaxation mode in the DWP,
$\omega_j$ the frequencies, $y_j$
the coordinates of the bath modes, and $\lambda_j$ the coupling of the
bath modes to the DWP.
 This potential energy term contains the full anharmonicity with respect
to the DWP mode. The other degrees of freedom are only considered up to their
harmonic contributions.  Of course, in order to obtain the full Hamiltonian of
the
system one has to add the kinetic energy of the relaxation mode and of the bath
modes.
For our simulation we chose a DWP with $V/k_{\rm B} \approx 300$~K.
Starting from one minimum we calculated all second
derivatives of the energy with respect to the positions of all 500 atoms,
yielding a $3N \times 3N$ dimensional matrix $A$.
Let $\hat{d}$ denote the unit vector in the $3N$-dimensional configuration
space connecting both minima of the DWP.
The matrix $A$ is diagonalized
in the space orthogonal to $\hat{d}$ , yielding eigenvalues
$a_{j}$ and eigenvectors $\hat{e}_{j}$. The corresponding
eigenfrequencies
$\omega_{j}$ are given by $\omega_j = \sqrt{a_j/m}$.
Three eigenvalues equal to zero reflect
the periodic boundary conditions. The coupling parameters $\lambda_{j}$
are calculated via $\lambda_j = \hat{d} A \hat{e}_{j}$.
It is worth noting that our vector $\hat{d}$ which is just a straight line  path
between the two minima of the DWP is not the reaction path between the two wells,
which would be curvilinear. However, the simulations in Refs. \CITE{Heu1,Heu2} have shown
that the curvature of the reaction path is not very large.

Later on it turns out that the relevant quantities describing the
coupling strength of the bath modes to the DWP is  the spectral densities
  \ba \label{M5}
  J(\om) =
  {\textstyle {2\over \hbar^2}}\sum_j c^{2}_j \delta(\om-\om_j)  
  \ea
where 
\begin{equation}
 c_j = \sqrt{\hbar/2m\omega_0}\, \sqrt{\hbar/2m\omega_j}\, \lambda_j\ .
\end{equation}
$J(\om)$ is plotted in Fig. 2. One can clearly see that bath modes exist for
all frequencies smaller than 400 K.
As a result,  DWPs relevant at the sound absorption peak
which only contain a few energy levels below the their barrier height $V$
are coupled to bath modes with frequencies as large as $V$.
The high-frequency modes are somewhat more localized
that the average modes and contain the dynamics of the order of 100 atoms. 
Such an
effect has been already observed in previous simulations.\cite{schober}

A closer inspection of the data shows that the coupling is strongly anharmonic.
This can be easily checked by repeating the above procedure calculating
the second derivatives at
different points along the reaction coordinate.
It turns out that the $\lambda_j^2$ and $\omega_j$ of the individual modes
strongly vary along $\hat{d}$ whereas the distribution function
$J(\omega)$ is insensitive. The reason for this is that $J(\omega)$ arises from
an ensemble average over a large number of oscillators whose 
parameter  variations  are statistically
independent  in  a first approximation.
However, one should keep in mind that due to the finite size of the simulation
box we do not obtain the low-frequency phonons in the Kelvin-regime
which are relevant for the
transitions between both minima at low temperatures. For those
phonons one expects that the anharmonic contributions are significantly
smaller. A more detailed discussion of the anharmonicities is beyond the scope
of the present paper.

\setcounter{equation}{0}
\section{The  Model}
We now construct a simple model which governs the relaxation dynamics
of the relevant DWPs. For reasons of simplicity we restrict ourselves
to only two pairs  of tunneling doublets below the barrier.

The phonons couple to the tunneling coordinate.
Their effect  is two-fold: They induce vibrational transitions
between the  eigenstates of the DWP
and destruct the tunneling coherence within each doublet.
Concerning the first effect we assume that  barrier penetrating
transitions are negligible.
The relevant matrix element between
the localized states reads
\begin{equation}
M = \langle 0\alpha |H|1 \alpha\rangle =
\sum_j c_j^{(v)} (b_j + b^{\dagger}_j)
\end{equation}
with phonon operators satisfying $[b_j,b_{j'}^{\dagger}] = \delta_{j,j'}$ and the coupling
constants $c_j^{(v)}\equiv c_j$ of each mode.
Here, the first index labels
the doublet and $\alpha = $L,~R the left and right well.
 The second effect of the phonon coupling is modeled by a diagonal coupling
which detunes $|n{\rm L}\rangle$
 against  $|n{\rm R}\rangle$ $(n,m=0,1)$.
 The corresponding matrix elements read
\begin{eqnarray}
e &=&  {\textstyle {1\over 2}} (\langle n{\rm R} |H|n {\rm R}\rangle
    - \langle n{\rm L} |H|n {\rm L}\rangle)\nonumber\\
  &=&
\sum_j c_j^{(t)} (b_j + b^{\dagger}_j)\
\end{eqnarray}
with (note that $ {\textstyle {1\over 2}} (\langle n{\rm R} |x|n {\rm R}\rangle
 - \langle n{\rm L} |x|n {\rm L}\rangle) \approx d$)
\ba
\frac{c_j^{(v)}}{c_j^{(t)}}
= \frac{\langle 0\alpha | x | 1 \alpha \rangle}{d} \equiv \kappa\ .
\ea
Corresponding to $c_j^{(v)}$ and $c_j^{(t)}$ we also define
the spectral densities $J_v(\omega)\equiv J(\omega)$ and 
$J_t(\omega)\equiv {1\over \kappa^2} J(\omega)$ where $J(\omega)$
has been defined in (\ref{M5}).

In case of our computer glass the factor  $\kappa$
can be estimated for symmetric DWPs as given by Eq. (\ref{pot}).
Since $\langle 0 \alpha | x |1 \alpha \rangle = \sqrt{\hbar/2m\omega_0}$
one directly obtains from Eq. (\ref{eqdwp1})
and Eq. (\ref{eq55})
$\kappa \approx 0.05$.
Hence, though the intra-well vibrational transition $M$
may be considered as a weak perturbation, the
diagonal coupling $e$, generally, may not. The main reason for this is
the ``polaron effect'', i.e.,  the slowing down
of the tunneling transition  associated
with the  shift of the oscillator coordinates into the other well.

As mentioned in Refs. \CITE{ME,PS} a suitable decomposition
of the Hamiltonian for second order perturbation theory
is achieved
 by the polaron transformation\cite{Po}
 ($\sigma_z ^{n} = |n{\rm L}\rangle\langle n{\rm L}|
 - |n{\rm R}\rangle\langle n{\rm R}|$) 
 \ba \label{M14}
 \e^S = \exp\left[- \sum_{n} \sigma_z ^{n}
 \sum_j \frac{c^{(t)}_j}{\hbar\omega_j} (b_j - b_j^{\dagger})\right]\ .
 \ea

We denote the energy separation of the pair of tunneling doublets by\cite{w}  
$\Delta E = \hbar\omega_{01}/2$ 
and the asymmetry by $\dd <
\hbar\omega_{01}$
(otherwise the states isolated in the lower
well will drop out of the dynamics and the doublets have to be redefined).
Now we may eventually define our model Hamiltonian.
 With the shift operator
 \ba \label{M15}
 B_\pm =
 \exp\left\{ \pm \sum_j \frac{2c^{(t)}_j}{\hbar\omega_j}
  (b_j - b_j^{\dagger})\right\}
 \ea
 and the tunneling frequency renormalized by a Debye-Waller factor
 $ \langle B_\pm\rangle_{\rm B} \equiv \e^{-W(T)}$ (see Appendix A),
\ba\label{RT}
{\it\widetilde{\Delta}}_n = \dd_n\, \e^{-W(T)}
\ea
it  reads
 in the localized  basis $|0{\rm L}\rangle$, $|0{\rm R}\rangle$, $|1{\rm
L}\rangle$,
  $|1{\rm R}\rangle$
 \ba\label{M16}
H &=& H_0 + H_{\rm B}  + H_{\rm int} \nn\\[0.3cm]
 &=& \pmatrix{ {1\over 2}\dd & {1\over 2}\hbar\widetilde{\dd}_0 & 0 & 0 \cr
             {1\over 2}\hbar\widetilde{\dd}_0 & -{1\over 2}\dd  & 0 & 0 \cr
 0 & 0 &  \hbar\om_{01} + {1\over 2}\dd & {1\over 2}\hbar\widetilde{\dd}_1 \cr
 0 & 0 &  {1\over 2}\hbar\widetilde{\dd}_1      & \hbar\om_{01} - {1\over 2}\dd
}
	\ + \ H_{\rm B}\nn\\[0.3cm]
  &+& \    \pmatrix{ 0 & {1\over 2}\hbar\dd_0  \delta B_- & M & 0 \cr
             {1\over 2}\hbar\dd_0 \delta B_+ & 0  & 0 & M \cr
	     M & 0 &  0 & {1\over 2}\hbar\dd_1 \delta B_- \cr
	  0 & M &  {1\over 2}\hbar\dd_1 \delta B_+ &  0 }
  \ea
  where $\delta B_\pm = B_\pm - \langle B_\pm\rangle_{\rm B}$, and
   $H_{\rm B} = \sum_j \hbar \om_j b_j^{\dagger}b_j$. Our goal is
to apply second order perturbation theory in $H_{\rm int}$.
In this approximation the intra-well dynamics is governed by the 
one-phonon transition between the vibrational levels, i.e., by 
the {\it Orbach process}.
In case of more than two doublets below $V$, we have to sum over
all pairs (see below).
We note in passing that the spectral density $J(\omega)$
only holds for transitions from level 0 to level 1. For
transitions changing the vibrational quantum number more than one
the transition matrix elements are only due to
the anharmonicity of the potential and
the interaction. This may lead to a somewhat decreased spectral density.
However, since we
are only interested in an order of magnitude estimation of the
coupling to the bath we do not analyse the influence of anharmonicies
on the coupling in greater detail.
Rather we state that for the observed weak intra-well coupling
(as compared to inter-well coupling due to $\kappa \ll 1$)
the Orbach process is indeed the dominant relaxation mechanism.

\setcounter{equation}{0}
\section{Dynamics}
 To determine the time evolution of the tunneling doublets we make use of
 Mori's projection operator technique \cite{Mori}.
We adopt throughout the paper the {\it fast vibrational relaxation limit}
which is relevant for most experimental situations.
Denoting the downwards vibrational transition  rate between the  two doublets
 by $\gam_{01}$ and the tunneling escape rate
 in the other well (inter-doublet transition rate)
by $\Gam_0$ and $\Gam_1$, this means that we assume
\ba\label{D4}
\Gam_0, \Gam_1 \ \ll\ \gam_{01}\ .
\ea
This guarantees local thermal equilibrium in each well.
The relaxation rates are defined in Appendix A.
 For  weak  coupling between the vibrational levels and
 in the temperature regime (\ref{TR}),   the upwards vibrational rate
 $\gam_{10} \equiv \gam_{01}\, \e^{-\beta\hbar\om_{01}}$
 is  typically much smaller than the level
 spacing
 \ba
\eps_n =  \sqrt{\dd^2/\hbar^2 + \widetilde{\dd}_n^2}\
\ea
of the two doublets [cf. Fig. 1].
  Furthermore, due to the strong increase
 of the tunneling matrix element near the barrier top, the ground state
 spacing is much less than the spacing near the barrier top. Hence we are
interested in the regime
  $\gam_{10} \ll\eps_0\ll\eps_1$  with no fixed relation between
$\gam_{01}$ and $\eps_0$, $\eps_1$.

 Relaxation dynamics is probed in long time or low frequency
 experiments where the time scale of the experiment is far off
 any resonant time scale of the system. Thus, any coherency has already
 decayed and the system shows pure decay.
 The experimentally
 accessible quantity is the {\it population difference} between the two wells.
The relevant operator measuring this quantity is
\ba \label{M9}
Q =  \pmatrix{ 1 & 0 & 0 & 0 \cr
               0 &  -1 & 0 & 0\cr
	       0 & 0  & 1 & 0 \cr
	       0 & 0 & 0 &   -1 }   \ .
\ea
 Its thermal expectation value reads
\ba \label{M10}
\langle Q \rangle = - \tanh(\beta\dd/2)\ .
\ea
 In linear response theory
 all information is contained in the symmetrized correlation
 function of the relevant operator
 \ba \label{M11}
C(t) &=& {\textstyle {1 \over 2}} \langle \delta Q(t)\; \delta Q\rangle
\  + \  t\leftrightarrow -t \nn\\
&=& {\textstyle {1\over 2}} \,
{\rm Tr}\, \Big[ \varrho\  \delta Q\, \e^{-i{\cal L}t}\, \delta Q \Big]
\  + \  t\leftrightarrow -t
\ea
where $\delta Q = Q - \langle Q \rangle$, and ${\cal L} = \hbar^{-1} [H,\ast]$
 is the Liouvillian.
Due to  local thermal equilibrium  in each well, the density matrix
  is given in the canonical form
   $\varrho = \exp(-\beta H_d)/{\rm Tr} \exp(-\beta H_d)$
   with a Hamiltonian $H_d$ which contains
  only the diagonal part of $H_0$ in the localized basis.
The  spectral function $C(\omega)  = (1/2)\int_{-\infty}^\infty C(t)
\e^{i\omega t} \d t$
can be directly measured  in neutron scattering experiments, or via the
dynamical susceptibility
in acoustic or dielectric experiments.

In Appendix B we calculate in second order
perturbation theory in $H_{\rm int}$
 the correlation function (\ref{M11})  by using Mori's
continued fraction representation for  the complex correlation function
$C(\lambda) =  \int_0^{\infty}  \e^{-\lambda t} C(t) \d t $ (Re($\lambda) >
0$).
After inverse Laplace transformation we find in the limit $\lambda \to  0$ [cf. Eq. B.5] 
our final result
 \ba \label{FG}
 C(t) \ = \ \langle (\delta Q)^2 \rangle\ \e^{- 2k t}
 \ea
 with $\langle (\delta Q)^2\rangle = {\rm sech}^2(\beta\dd/2)$ and the escape
rate
 \ba \label{DK2}
 k =  k_t + k_v  \ .
 \ea

 The tunneling rate $k_t$ is given by the usual small polaron expression
 averaged over the thermal level occupation
\ba\label{D5}
k_t &=& {\textstyle{1\over 2}} \langle \Gam_0 + \Gam_1 \rangle_\beta\nn\\
&=& {\textstyle {1\over 4\hbar^2}}\int_{-\infty}^\infty \d t \
\langle \widetilde{\dd}_0^2 + \widetilde{\dd}_1^2\rangle_\beta\,
 \cos (\dd t/\hbar)\,
\Big( \cos[g_1(t)]\, \e^{g_2(t)} - 1 \Big)
\ea
where
\ba
\langle \widetilde{\dd}_0^2 \rangle_\beta &=&
 \frac{\widetilde{\dd}_0^2}{1 + \e^{-\beta\hbar\omega_{01}}}\\
 \langle \widetilde{\dd}_1^2 \rangle_\beta &=&
 \frac{\widetilde{\dd}_1^2}{\e^{\beta\hbar\omega_{01}} + 1}\ .
 \ea
 In the  fast vibrational relaxation (\ref{D4})
 the effect of the vibrational transitions
 enters $k_t$ only through the thermal occupation  number of  the localized
states.

 The vibrational relaxation  rate
 \ba\label{r1}
  k_v \ =\  {1\over 2}\,\frac{\widetilde{\dd}_1^2}
        {\eps_1^2 + \gam_{01}^2}
            \  \frac{\gam_{01}\,
            }{\e^{\beta\hbar\om_{01}} + 1}\
 \ea
    is characterized by the ratio of
   the doublet splitting $\eps_1$ and the equilibration rate
 $\gam_{01}$ between  vibrational levels.
This type of expression for $k_v$ has first been proposed
by Sussmann\cite{S} on ground of a simple wave function argument.
 For $\dd = 0$ it is 
  identical to the transition rate between E and A symmetric
  states in a rotational tunneling problem.\cite{W,Heu}

  \setcounter{equation}{0}
  \section{Discussion}
\subsection{Analogy to the spatial- and energy-diffusion-controlled limit}
Our results can be summarized in the following physical picture.
There exist two relaxation mechanism
for a particle in the energetically instable well :
(1) Incoherent tunneling with rate $k_t$
within the thermally occupied tunneling doublets, or (2)
vibrational transition to a non-thermally occupied doublet
of higher energy, coherent tunneling within the doublet
and vibrational decay to the lower well which results
in a rate $k_v$. This interpretation also follows nicely
from the projection operator method used in Appendix B. 
Relevant degrees of freedom are the diagonal terms in $H_0$, i.e.,
the population in the left and right well.  In  a projection
operator formalism,  relevant degrees of freedom (projected out by 
$\cal{P}$ at an earlier time $t' < t$) couple via an interaction ($\cal{L}$) to irrelevant 
degrees of freedom (projected out by 
$\cal{Q}$), which subsequently evolve in time ($\cal{QLQ}$) and, 
due to a second interaction ($\cal{L}$), acquire relevancy again, thus
influencing the evolution of the relevant degrees of freedom at the present time $t$.
Eqs. (B.6) and (B.7) are just the Laplace transform of the kernel describing this process.
Now, $k_t$ propagates that part of the irrelevant degrees of freedom which destruct
coherence of the relevant degrees of freedom, i.e., $\dd_n \delta B_\pm$, whereas
$k_v$ propagates that part which maintains coherence, i.e., $\widetilde{\dd}_n$.

Process (1) has been discussed
by Parris and Silbey \cite{PS} and Meyer and Ernst \cite{ME}.
This process always prevails  if $\gamma_{01} \gg \epsilon_1$
because then the particle has no time for tunneling before decaying
from the upper doublet.
In this case the preexponential factor  of $k_v$
   \ba \label{DTT5}
   f_{01} =   {1\over 2}\,\frac{\widetilde{\dd}_1^2\, \gam_{01}}
   {\eps_1^2 + \gam_{01}^2}
    \ea
   is approximately $f_{01} \approx
\widetilde{\dd}_1^2/ 2\gam_{01}$.  This is
the quantum analog of the {\it spatial-diffusion-controlled} limit
in the classical Kramers' problem.\cite{K}
The escape to the other well is hindered by
the damping of the tunneling motion,  i.e., $k\approx k_t$.
The tunneling escape rate $k_t$
is given by the Boltzmann-weighted
average of the tunneling equilibration rate within each tunneling doublet
at given vibrational level.
This equilibration is hindered by the distortion of the lattice in the
tunneling process, i.e., by the polaron effect.
The suppression
of the vibrational rate for $\gam_{01} \gg \eps_1$ can be thought off as a
{\it phonon bottleneck effect}.\cite{S} The physical reason for the
suppression of $k_v$    is that
phonons resonating between the upper and the lower
level of two different  tunneling doublets are
not distinct because the level width exceeds the doublet spacing.
As a result one cannot add up their rates. Instead
they can interfere resulting in a suppression of $k_v$.

Process (2) has first been proposed by Sussmann\cite{S} 
on ground of a simple wave function argument.
It prevails provided that  $(i)$ the vibrational transition to the upper
doublet is faster than the direct decay via incoherent tunneling,
\begin{equation} \label{p}
\gam_{10} \ \equiv \
 \gam_{01}\ \e^{-\frac{\hbar\om_{01}}{k_{\rm B}T}}\  >\
\Gam_0\ ,
\end{equation}
and, necessarily,  if $(ii)$ the decay of the excited state is slower
than the tunneling process, $\gam_{01} < \eps_1$. In this case
the particle has sufficient time for oscillating   coherently fro and back
in the upper level, and, therefore, a maximal,  i.e., 50\% probability
for decaying in the other well.
As a result the energy transport to  the barrier top  becomes the
rate limiting process, i.e., $k\approx k_v$.
This is the  situation analogous  to
the {\it energy-diffusion-controlled} limit in the classical Kramers' problem.
It should be noted that the  quantity
  which determines the frequency of coherent oscillations
is the doublet spacing
$\epsilon_1 = \sqrt{\widetilde{\dd}_1^2 + \dd^2/\hbar^2}$ and not the tunneling
frequency
$\widetilde{\dd}_1$.
Hence, the asymmetry acts in suppressing the  bottleneck effect.

Clearly  process (2) becomes increasingly likely with increasing
temperature. Eq. (\ref{p}) defines a
 transition temperature $T^*$.
Furthermore a large $\epsilon_1$
is needed. Hence,  upper doublets with  energies  of $O(V)$ are preferable.

\subsection{Apparent Arrhenius behavior}
Now we present a derivation of an {\it apparent}
Arrhenius behavior at higher temperatures with an activation energy
$O(V)$.
 We still consider the situation where  the eigenvalues are discrete
and not dense at $E\approx V$.
Our previous results are now generalized  to many pairs
 by noting that the total rate is  the sum over
  all pair rates   weighted by the Boltzmann occupation
  factor of the initial state. We denote the (downwards) equilibration rate
  between the $m$th and the $n$th vibrational state ($n < m$) by $\gam_{nm}$.
 This provides the rate
 \ba \label{DKK2}
 k &=& k_t + k_v \nn\\[0.25cm]
 &=& \left(\, {\textstyle {1\over 2}}
 \sum_n \Gam_n \, \e^{-\beta E_n}\ + \
\sum_{n<m}
   \, \ f_{nm}\, /  [\e^{\beta E_m} + \e^{\beta E_n}]\right)\,  \left/ \,
 \sum_n \e^{-\beta E_n} \right. .
 \ea
 with
 \ba
 f_{nm} = \frac{1}{2}
 \frac{\widetilde{\dd}^2_m \gamma_{nm}}{\epsilon_m^2 + \gamma_{nm}^2}
 \ea
 For $(n=0,m=1)$ the term for $k_v$ boils down to Eq. (\ref{r1}).
 Now we consider the relaxation of a single DWP with given parameters.
 As discussed above for $T > T^*$ the relaxation is mainly determined
 by $k_v$. In order to obtain the temperature dependence one has
 to check by which doublets $(n,m)$ this term is dominated. As a first
 approximation we neglect the dependence of $\gamma_{nm}$ on $n$ and $m$,
 hence introducing a single value of $\gamma$.
 Then $f_{nm}$ only depends on $m$.
 This may be a poor approximation. However, since other quantities
 like the tunneling matrix element or the Boltzmann factor exponentially
 depend on the indices this approximation is sufficient for our
 present purposes.

 We now ask  by which value of $m$
 the term $f_{nm}$ is dominated.
 We first note that only doublets with $\epsilon_m > \gamma$
can avoid the bottleneck effect, and that, even among those,
doublets with $\widetilde{\dd}_m/\dd \ll 1$ are unimportant.
 Evidently, $f_{nm}$  cannot grow beyond
 $\gamma/2$ ---
the strict maximum is only reached for $\widetilde{\dd}_m^2
 \rightarrow \infty$.
However, already for
 \ba
 \label{eqcrit}
 \widetilde{\dd}_m^2 = \dd^2/\hbar^2 + \gamma^2
 \ea
 the value $\gamma/4$ is reached.
 Hence,  Eq. (\ref{eqcrit})  is an appropriate criterion that
 $f_{nm}$ is close to its maximum value.
One can estimate from the numerical data
that the average value of
$\pi J(\omega) k_{\rm B}/\hbar$ is of the order of
30$~K$. Having in mind that
this estimation may be somewhat too large (see Sec. 3)
we may estimate $\gamma \hbar / k_{\rm B} = O(10$ K).
This is somewhat smaller but of the same order as the numerical value
we have found for $\omega_0$. Hence,
at least  for asymmetric DWPs,  criterion (\ref{eqcrit})
can hardly be fulfilled for real systems. However,
doublets which come closest to (\ref{eqcrit}) will still dominate
$f_{nm}$. Evidently, these are
 levels $m$ close to the barrier.
This is related to  the exponential
dependence of the tunneling matrix element on the
energy of the corresponding level.
Only close to the barrier the maximum value
$\widetilde{\dd} \approx 2 \omega_0/\pi$ 
 is reached.\cite{com}

Additionally taking into account the Boltzmann factor one can
see that all levels $n$ contribute as long as $\exp(\beta E_m)
\gg \exp(\beta E_n)$. For $\beta E_m >> 1$ this implies that
most levels $n$ indeed contribute.

Summarizing the arguments presented above we obtain
 \begin{equation}\label{kl}
 k_v \ \approx \ \tau^{-1}_0\, \ \e^{-\beta\Delta U}
 \end{equation}
with  an activation energy $\Delta U = O(V)$
and a prefactor  $\tau_0^{-1}$ which is of the order
$\gamma/4 \times m$. The latter factor results from the
summation of most of the states below the $m$-th level.
Since our numerical results indicate that
$\gamma$ is somewhat smaller but close to $\omega_0$
we may write $\tau^{-1}_0 \approx O(\omega_0/2\pi)$.
As a surprising result,
for sufficiently high temperatures, $T>T^*$ such that $k_v\gg k_t$,
the relaxation rate between the two wells
is an effective Arrhenius rate where both the prefactor
and the activation energy are of the same
order of magnitude as in the classical limit.

\section{Summary and Conclusion}
The intension of this work was
to investigate the remarkable success of fitting
relaxation data of disordered solids at higher temperatures
by a simple Arrhenius rate despite the fact the underlying
classical picture  might not  be justified automatically.

Due to the lack of knowledge of microscopic DWPs in disordered solids
we have used numerical simulations on a computer glass to
characterize  the DWP and the bath modes before solving the dynamics
for this particular situation. It should be noted that these
simulations have proven to be reliable by  reproducing  characteristic
low-temperature
properties of glasses.\cite{Heu1,Heu2} Hence,  the parameter regime
for which we have solved the dynamics can be considered
as ``typical'' for structurally disordered solids.
The simulations have taught us that DWPs relevant at the
sound absorption peak, i.e., those with large  $V$,
only contain a few energy levels below their barrier height $V$  and are
coupled to bath modes with frequencies as large as $V$.

To resolve the paradox to  the apparent classical behavior
we have constructed a simple model of two  tunneling doublets
which are coupled by intra-well vibrational transitions.
Independently of how closely this model reflects the true
physical situation in detail, it has provided a simple physical picture
how  even in
this deep quantum mechanical regime
an apparent Arrhenius behavior might emerge.
Solving our model by  standard approximations we could confirm the
suggestion of  Sussmann \cite{S}: at high temperatures
 the vibrational transition
to a doublet near the barrier top, followed by coherent tunneling between
the wells and vibrational decay to the bottom becomes a faster
relaxation mechanism than direct decay via incoherent tunneling.
Clearly the derivation  which eventually led to the Arrhenius rate (\ref{kl})
is very scetchy.
In general there will be corrections to the term satisfying
(\ref{eqcrit}) which are in the present formulation difficult
to quantify. Furthermore, if more than two doublets
are involved, it is difficult to define quantitatively the transition
temperature $T^*$ which separates the tunneling from the
vibrational dominated regime.
Hence, generally, both terms $k_t$ and $k_v$  may contribute
and the Arrhenius rate has to be viewed as an idealization.

However, we believe that  the picture drawn  provides
a physical intuition on the remarkable fact
that a simple Arrhenius rate is so successfull in explaining
relaxation data even just above temperatures where
the quantum mechanical tunneling effect had dominated the
relaxation dynamics.
In addition to this it is reasonable to  expected that in disordered
solids like glasses the detailed form of the barrier distribution
function has a much greater impact than possible
corrections to the Arrhenius rate. Hence, the customary
application of this rate for practical purposes can be
 justified {\it a posteriori} by the present picture
even in the deep quantum regime.

\acknowledgements
Helpful discussion with   H. Horner, R. Meyer, D. R. Reichman
and R. J. Silbey are gratefully
acknowledged.  This work has been financially supported by the NSF
and the Alexander von Humboldt foundation.

 %\newpage

\setcounter{equation}{0}
\appendix

\renewcommand{\theequation}{\Alph{section}.\arabic{equation}}

%\section*{Appendix}

\section{ Bath correlation functions and relaxation rates}
 Assuming that the bath is not disturbed by the DWP we can calculate
 bath correlation functions with $H_{\rm B}$ and
$\varrho_{\rm B} = \exp(-\beta H_{\rm B})/Z_{\rm B}$.
Defining
\ba \label{M17}
g(t) &=& {\textstyle {\hbar^2\over 2}} \int_0^\infty \d\omega\,  J_v(\omega) \,
              \left(n(\omega) \e^{i\omega t} +
	       \big(n(\omega)+1\big) \e^{-i\omega t} \right)\\
g_1(t) &=& 2 \int_0^\infty \d\omega \frac{J_t(\omega)}{\omega^2} \sin(\omega
t)\\
g_2(t) &=& 2\int_0^\infty \d\omega \frac{J_t(\omega)}{\omega^2}
                  \coth(\beta\hbar\omega/2) \cos(\omega t) \\
W(T) &=& {\textstyle {1\over 2}} g_2(0)
\ea
the bath correlation functions read
\ba \label{M18}
\langle B_+(t) B_-\rangle_{\rm B} &=& \langle B_-(t) B_+\rangle_{\rm B} =
\langle B_+ B_-(t)\rangle^\ast_{\rm B} =
\langle B_- B_+(t)\rangle^\ast_{\rm B}\nn\\
&=& \exp\Big[g_2(t) - ig_1(t)\Big] \,\e^{-2W(T)}\\
\langle M(t)  M \rangle_{\rm B} &=& g(t)
\ea
Furthermore, we need the following modification of the vibrational
relaxation function
\ba \label{M19}
g_\pm(\omega_0,t)\  =\  \e^{i\omega_{0}t}\
{\textstyle{\hbar^2\over 2}} \int_0^\infty \d\omega\,  J_v(\omega) \,
              \left(n(\omega) \e^{\pm i\omega t} +
			  \big(n(\omega)+1\big) \e^{\mp i\omega t} \right)\ .
\ea
In the following two kind  of relaxation rates are needed.
First,  the transition rate {\it within} the $n$th doublet (inter-well transition)
\ba \label{D1}
\Gam_n \ = \ {\textstyle {1\over 2\hbar^2}}\int_{-\infty}^\infty \d t \
\widetilde{\dd}_n^2\, \cos (\dd t/\hbar)\,
\Big( \cos[g_1(t)]\, \e^{g_2(t)} - 1 \Big)\ .
\ea
This expression is well-known from small polaron theory.

The second rate is the transition rate {\it between} the   doublets (intra-well transition)
\ba \label{D2}
\gamma_{10} &=& {\textstyle \frac{1}{2\hbar^2}} \int_{-\infty}^\infty \d t
\Big( g_-(\omega_{01},t) + g_+(-\omega_{01},t)\Big) \\
\gamma_{01} &=& {\textstyle \frac{1}{2\hbar^2}} \int_{-\infty}^\infty \d t
\Big( g_-(-\omega_{01},t) + g_+(\omega_{01},t)\Big) \ .
\ea
$\gamma_{10}$ is the transition rate from the zeroth  to the first
vibrational level and given by
\ba\label{1ph}
\gamma_{10} = \pi J_v(\omega_{01}) n(\omega_{01})\ .
\ea
Here, $n(\omega_{01}) \equiv [\exp(\beta\hbar\omega_{01}) - 1]^{-1}$
is the Bose factor.
  $\gamma_{01}$ and $\gamma_{10}$ satisfy the principle of detailed balance
$\gamma_{10} = \gamma_{01} \,\e^{-\beta\hbar\omega_{01}} $.

\setcounter{equation}{0}
\section {Formulation of the dynamics in the
Mori scheme}

If we define the symmetrized scalar product $(A|B):= {1\over 2} \langle
A{^\dagger} B + B A^{\dagger}\rangle$ we can write
 the complex correlation function
$C(\lambda) =  \int_0^{\infty}  \e^{-\lambda t} C(t) \d t $ (Re($\lambda) > 0$)
 as a resolvent matrix element
\ba \label{DT1}
C(\lambda) = (\delta Q| [\lambda  + i {\cal L}]^{-1}| \delta Q)\  .
\ea
We define the projector
${\cal P} = |\delta Q)\eta^{-1}(\delta Q| = {\cal I} - {\cal Q}$
and its complement ${\cal Q}$
with the normalization
$\eta \ = \  (\delta Q|\delta Q) \ \equiv\ {\rm sech}^2(\beta\dd/2)$.
 The resolvent identity
\ba
[\lambda + i {\cal L_{PP}} +
{\cal L_{PQ}} [\lambda  + i {\cal L_{QQ}}]^{-1}{\cal L_{QP}}]\,
{\cal P}[\lambda  + i {\cal L}]^{-1}{\cal P} = {\cal P}
\ea
where ${\cal L_{AB}} =  {\cal ALB}$ with ${\cal A,B}\in\{{\cal P,Q}\}$
defines a relaxation kernel $k(\lambda)$.
With $(\delta Q|{\cal L}|\delta Q) = 0$
we find
\ba\label{D3}
C(\lambda) = \frac{\eta}{\lambda +  2k(\lambda)}
\ea
where the relaxation kernel is given by
\ba\label{D3a}
 2k(\lambda) = (\delta Q| {\cal L_{PQ}}
[\lambda  + i {\cal L_{QQ}}]^{-1}|{\cal L_{QP}} \delta Q)\, \eta^{-1}\ .
\ea
For long times the transition rate $k$  between the left and right well is
given
 by the limit $\lambda \to 0$
\ba
k \ = \ \lim_{\lambda\to 0} k(\lambda)\ .
\ea
To calculate this function,
we first  separate $k(\lambda)$ into two terms
$k(\lambda) = k_t(\lambda) + k_v(\lambda)$
with
\ba
2 k_t(\lambda) &=&
(\delta Q| {\cal L_{PQ}}
 [\lambda  + i {\cal L_{QQ}}]^{-1}|\delta  R)\, \eta^{-1} \\
2 k_v(\lambda) &=&
(\delta Q| {\cal L_{PQ}}
 [\lambda  + i {\cal L_{QQ}}]^{-1}|\widetilde{R})\, \eta^{-1}
\ea
where $\delta R$ contains all matrix elements of ${\cal L_{QP}} \delta Q$
arising from $H_{\rm int}^{(t)}$ and $\widetilde{R}$ contains all matrix
elements
of ${\cal L_{QP}} \delta Q$ arising from the non-diagonal part
of $H_0$. Note that $H_{\rm int}^{(v)}$ does not contribute to ${\cal L_{QP}}
\delta Q$.

We now apply lowest order perturbation theory in $H_{\rm int} =
H_{\rm int}^{(t)} + H_{\rm int}^{(v)}$,  i.e.,
 in $\dd_n\delta B_\pm$ and $M$.
 Since terms which are proportional to the
product of these two perturbations are of higher order, we neglect
the vibrational coupling $M$ in the
tunneling part $k_t$,  and the tunneling coupling $\dd_n\delta B_\pm$
in the vibrational part $k_v$.

It is  a standard result
of spin-boson literature that for a single tunneling doublet
the  tunneling transition rate defined by
$k_t \equiv    k_t(\lambda = 0)$ is given by expression
(\ref{D1}) in second order perturbation theory
in the tunneling matrix element.
Repeating the same kind of calculation for the pair of doublets
one easily finds (\ref{D5}) in case of fast vibrational relaxation (\ref{D4}).

  The calculation of the vibrational rate is essentially identical
  to the calculation of the transition  rate between E and A symmetric states
  in a XH$_2$ rotational tunneling problem.
Let us introduce the notation that  $E_{nm}$ denotes a matrix  with
a one in the $n$-th row and the $m$-th column and zeros elsewhere.
With this, we separate $\widetilde{R}$ into two part
$ \widetilde{R}  =  \widetilde{\dd}_0 R_0  +   \widetilde{\dd}_1 R_1$
 with $R_0 = E_{21} - E_{12}$ and $R_1 = E_{43} - E_{34}$.
 Inserting this and the projector
\ba
{\cal P}' = \sum_{\alpha = 0,1} |R_\alpha)\eta_\alpha^{-1}(R_\alpha|
\ea
with $\eta_0 = [1 + \e^{-\beta\hbar\omega_{01}}]^{-1}$ and
$\eta_1 = [\e^{\beta\hbar\omega_{01}} + 1]^{-1}$ into $k_v(\lambda)$
 and noting that
 $(\delta Q| {\cal L}| R_0) \eta_0^{-1} \eta^{-1} = \widetilde{\dd}_0$  and
$(\delta Q| {\cal L}| R_1) \eta_1^{-1} \eta^{-1} = \widetilde{\dd}_1$
one finds that the  vibrational part of the relaxation kernel becomes
\ba
 k_v(\lambda) \approx
 \widetilde{\dd}_1^2\, G_{11}(\lambda)
\ea
with
\ba
G_{11}(\lambda) = (R_1| [\lambda +  i{\cal L_{QQ}}]^{-1}| R_1)\ .
\ea
Because of  $\widetilde{\dd}_1 \gg \widetilde{\dd}_0$,  we have neglect
all terms $\propto \widetilde{\dd}_0$.
The calculation of this function in second order perturbation theory  in
the vibrational coupling $M$ allows the replacement
${\cal L_{QQ}} \to {\cal L}'  = \hbar^{-1} [H_0 + H_{\rm int}^{(v)},\ast]$
where
$H_0 + H_{\rm int}^{(v)}$ have been defined in (\ref{M16}).
The problem is now essentially idential to the calculation of the scattering
function
in a XH$_2$ rotational tunneling system. Details of this can be found in Refs.
\CITE{W,Heu}.
Following essentially their lines one finds with the definition
$k_v \equiv k_v(\lambda = 0)$ the result (\ref{r1}).

% \newpage

%\begin{thebibliography}{99}

%\end{thebibliography}

\subsection*{FIGURE CAPTIONS}

\begin{itemize}
\item[FIG. 1:] Schematic representation of a biased
               double-well potential with $V\approx 2 \hbar\omega_0$.
               The ground state and first excited
               state tunneling doublets are separated by
               an energy $\Delta E \approx \hbar\omega_0$.
\item[FIG. 2:] Spectral density $J_v(\omega) \equiv J(\omega)$ 
               for the NiP Lennard-Jones computer glass.
\end{itemize}


\begin{references}


\bibitem{Na} S. Nakaoka, T. Terao, F. Imashiro, N. Hrota, and
             S. Hayashi, J. Chem. Phys. {\bf 79}, 4694 (1983).

\bibitem{Sk} J. L. Skinner and H. P. Trommsdorff,
             J. Chem. Phys. {\bf 89}, 897 (1988).

\bibitem{ME} R. Meyer and  R. R. Ernst,  J. Chem. Phys. {\bf 93}, 5518 (1990).

\bibitem{SS} A. Suarez and  R. J. Silbey, J. Chem. Phys. {\bf 95}, 4201 (1991).

\bibitem{HH} A. Heuer and U. Haeberlen,  J. Chem. Phys. {\bf 95}, 4201 (1991).

\bibitem{Mak} V. A. Benderskii, V. I. Goldanskii, and D. E. Makarov,
              Phys. Rep. {\bf 233}, 195 (1993).


\bibitem{Hunk} D. Tielb\"urger, R. Merz, R. Ehrenfels, and S. Hunklinger;
               Phys. Rev. B {\bf 45}, 2750 (1992).

\bibitem{Rau} S. Rau, C. Enss, S. Hunklinger, P. Neu, and A. W\"urger,
              Phys. Rev. B {\bf 52}, 7179 (1995).

\bibitem{Legg} A. J. Leggett, S. Chakravarty. A. T. Dorsey, M. P. A. Fisher,
               A. Garg, and W. Zwerger, Rev. Mod. Phys. {\bf 59}, 1 (1987).


\bibitem{Weisb}    U. Weiss, {\it Quantum Dissipative Systems},
                Series in Modern Condensed Matter Physics Vol. 2
                (World Scientific, Singapore, 1993).

\bibitem{Var} R. J. Silbey and R. A. Harris, J. Chem. Phys. {\bf 80},
               2615 (1984).

\bibitem{Mode} R. Beck, W. G\"otze, and P. Prelovsek,
               Phys. Rev. A {\bf 20}, 1140 (1979);
               W. Zwerger, Z. Phys. B {\bf 53}, 53 (1983);
               W. G\"otze and G. M. Vujicic,
               Phys. Rev. B {\bf 38}, 87 (1988);
               P. Neu and A. W\"urger, Z. Phys. B {\bf 95}, 385 (1994).

\bibitem{PS}  P. E. Parris and  R. J. Silbey,
              J. Chem. Phy. {\bf 83}, 5619 (1985);
              D. R. Reichman and  R. J. Silbey,
              J. Phys. Chem. {\bf 99}, 2777 (1995).

\bibitem{Phi87}  W. A. Phillips, Rep. Prog. Phys. {\bf 50}, 1657 (1987).

\bibitem{Heu1}  A. Heuer, R. J. Silbey, Phys. Rev. Lett. {\bf 70}, 3911 (1993).

\bibitem{Heu2} A. Heuer, R. J. Silbey, Phys. Rev. B {\bf 53}, 609 (1996).

\bibitem{schober}  H. R. Schober, C. Oligschleger,  and B. B. Laird,
                   J. of Non-Crystalline Solids 156-158, 965 (1993).

\bibitem{Po} The resulting polaron
 factor for the vibrational coupling $M$ can be neglected, since
 there is almost no lattice distortion in this transition -- the tranfer
 distance is essentially zero in this case \cite{ME}. Furthermore, the
resulting energy
 shift $\sum_j c^{(t)2}_j/\hbar \omega_j$
 for the diagonal elements can be neglected  since this
 shift is almost the same for all energy levels.

\bibitem{w} For this simplified model $\omega_{01}$ is identical to the
zero-point
frequency $\omega_0$.
In the original DWP, however, the  energy
difference between doublets constituting a pair
may exceed several $\omega_0$ because the coupled
doublets  are not necessarily adjacent.

\bibitem{K} In the classical regime, Kramers has determined the effect of
friction
on the preexponential factor in (\ref{1}) for a particle subject to frequency
independent damping $\gam$ and Gaussian noise
[for a review see Refs. \CITE{Weisb,Ha}]. In the strong
damping limit $\gam\gg\om_b$, where $\om_b$ is the barrier
frequency, the prefactor is reduced as compared to $\om_0/2\pi$
by a factor $\om_b/\gam$ because of diffusive barrier recrossings.
This has been termed {\it spatial-diffision-controlled limit}.
On the other hand, for weak friction,
 typically $\gam \< \om_b k_{\rm B}T/V$,
 the influence of the heat bath is not strong enough to maintain
 thermal equilibrium at the top of the barrier. In this case
 the energy flow to the barrier top works as a bottleneck
 which reduces the prefactor as compared to $\om_0/2\pi$
approximately by  $\gam V/\om_b k_{\rm B} T$. This
situation has been termed {\it energy-diffusion-controlled limit}. Since this
seminal work,  quantum  corrections to the classical escape rate
 have been elaborated showing explicitly the crossover
 between the quantum and the classical regime
[see for instance Refs. \CITE{Mak,Weisb,Ha,HAN} and references therein].


\bibitem{Ha} P. H\"anggi, P. Talkner and M. Borkovec, Rev. Mod. Phys.
            {\bf 62}, 251 (1990).

\bibitem{HAN} P. H\"anggi and P. Fleming, {\it Activated barrier crossing:
          Applications in physics, chemistry and biology},
                (World Scientific, Singapore, 1993).


\bibitem{S} J. A. Sussmann, Ann. Phys. {\bf 6}, 135 (1971).

\bibitem{W} A. W\"urger, J. Phys.: Condens. Matter {\bf 1}, 6901 (1989).

\bibitem{Heu}  A. Heuer,  Z. Phys. B {\bf 88}, 39 (1992).

\bibitem{Mori} E. Fick and G. Sauermann,
                  {\it The Quantum Statistics of Dynamical Processes},
                 Springer Series in Solid-State Sciences {\bf  86}
                 (Springer,  Berlin Heidelberg New York, 1990).


\bibitem{com}   This
follows from a WKB relation $\dd(E) \ =\ \frac{2\om_0}{\pi}\
  \exp(-{1\over\hbar}\int_{-a}^a \sqrt{2\mu(V(x)-E)}\, \d x)$. See
A. Das, {\it Quantum Field Theory, A Path Integral Approach},
                Lecture Notes in  Physics Vol. 52, Chap. 7.4
                (World Scientific, Singapore, 1993).



\end{references}
\end{document}